\documentclass[10pt]{article}
\usepackage{theorem}
\usepackage[mathscr]{eucal}
\usepackage{amssymb}
\newtheorem{prop}{}[section]
{\theorembodyfont{\upshape} }
\begin{document}
\hyphenation{uni-que-ness}
\newcommand{\rr}[2]{{#1}^{(#2)}}
\newcommand{\qq}[2]{{#1}^{[#2]}}
\newcommand{\boma}[1]{{\mbox{\boldmath $#1$} }}
\newcommand{\co}[2]{ {(#1)_{#2} \over #2!} }
\newcommand{\cp}[2]{ {(#1)_{#2} \over (#2)!} }
\newcommand{\Int}[1]{{#1}^{\circ}}
\def\Treves{Tr\`eves~}
\def\gen{\boma{\xi}}
\def\vvv{\boma{v}}
\def\fun{h}
\def\tfun{\tilde{\fun}}
\def\tF{\tilde{F}}
\def\tpsi{\tilde{\psi}}
\def\tf{\tilde{f}}
\def\tL{\tilde{L}}
\def\be{\beta}
\def\bei{\beta^{\dag}}
\def\ber{\beta^{\ddag}}
\def\leqs{\leqslant}
\def\geqs{\geqslant}
\def\lg{{\scriptscriptstyle{\barray{c} \vspace{-0.15cm} \leqs \\ \vspace{-0.0cm} \geqs \farray}}}
\def\supinf{{\scriptscriptstyle{\barray{c} \vspace{-0.05cm} \sup \\ \vspace{-0.0cm} \inf \farray}}}
\def\piu{{\raisebox{0.05cm}{\mbox{$\scriptscriptstyle +$}}}}
\def\meno{{\scriptscriptstyle -}}
\def\vers{\mbox{vers}}
\def\ct{\complessi^{\times}}
\def\cz{\complessi_0}
\def\ctt{\complessi^{* \times}}
\def\ctz{\ct_{\, 0}}
\def\cttz{\ctt_{~0}}
\def\Re{\mathscr R}
\def\Im{\mbox{Im}}
\def\Ro{\mathfrak R}
\def\Se{\mathscr S}
\def\Ti{\mathfrak T}
\def\implica{\Longrightarrow}
\def\sign{\mbox{sign}}
\def\di{\lambda}
\def\er{r}
\def\oa{$\overline{\mbox{a}}$}
\def\Ei{\mbox{Ei}}
\def\tu{t_1}
\def\td{t_2}
\def\tm{t_m}
\def\tM{t_M}
\def\n{\nu}
\def\tv{t_{\v}}
\def\Lp{L^{>}_{\v}}
\def\Lm{L^{<}_{\v}}
\def\ffi{\varphi}
\def\dfi{\delta \varphi}
\def\fip{\dot{\varphi}}
\def\dfip{\delta \dot{\varphi}}
\def\ES{{\mathscr S}}
\def\gi{{\tt g}}
\def\om{{\omega}}
\def\J{{\mathscr J}}
\def\H{{\mathscr H}}
\def\K{{\mathscr K}}
\def\Kp{{\mathscr K}'}
\def\kp{k'}
\def\scrscr{\scriptscriptstyle}
\def\scr{\scriptstyle}
\def\dd{\displaystyle}
\def\SB{\mathscr{B}}
\def\B{ B_{\mbox{\scriptsize{\textbf{C}}}} }
\def\Bc{ \overline{B}_{\mbox{\scriptsize{\textbf{C}}}} }
\def\ppartial{\overline{\partial}}
\def\d{\hat{d}}
\def\TT{{\mathcal T}}
\def\G{ {\textbf G} }
\def\Hinf{ H^{\infty}(\reali^d, \complessi) }
\def\Hn{ H^{n}(\reali^d, \complessi) }
\def\Hm{ H^{m}(\reali^d, \complessi) }
\def\Ha{ H^{\d}(\reali^d, \complessi) }
\def\Ld{L^{2}(\reali^d, \complessi)}
\def\Lpi{L^{p}(\reali^d, \complessi)}
\def\Lq{L^{q}(\reali^d, \complessi)}
\def\Lr{L^{r}(\reali^d, \complessi)}
\def\Knb{K^{best}_n}
\def\k{\mbox{{\tt k}}}
\def\x{\mbox{{\tt x}}}
\def\D{\mbox{{\tt D}}}
\def\g{ {\textbf g} }
\def\KdV{\scriptscriptstyle{KdV}}
\def\mKdV{\scriptscriptstyle{mKdV}}
\def\ccc{\,\makebox[0.12cm]{{$\boma{\circ} \hskip -0.190cm \circ$}}\,}
\def\QQQ{\boma{Q}}
\def\XXX{\boma{X}}
\def\HHH{\boma{H}}
\def\iint{\,\makebox[0.12cm]{${\boma{\int \hskip -0.28cm \int}}$}\,}
\def\hhh{\boma{h}}
\def\fff{\boma{f}}
\def\ZZZ{\mathfrak Z}
\def\AAA{\boma{\mathfrak F}}
\def\BBB{\mathfrak B}
\def\MMM{\boma{M}}
\def\RRR{\boma{R}}
\def\FFF{\boma{F}}
\def\GGG{\boma{G}}
\def\PPP{\boma{P}}
\def\DDD{\boma{\partial}}
\def\gr{\mbox{graph}~}
\def\Q{$\mbox{Q}_a$~}
\def\PZ{$\mbox{P}^{0}_a$~}
\def\PZAL{$\mbox{P}^{0}_\alpha$~}
\def\PL{$\mbox{P}^{1/2}_a$~}
\def\PU{$\mbox{P}^{1}_a$~}
\def\PK{$\mbox{P}^{k}_a$~}
\def\PKU{$\mbox{P}^{k+1}_a$~}
\def\PI{$\mbox{P}^{i}_a$~}
\def\Pell{$\mbox{P}^{\ell}_a$~}
\def\PTM{$\mbox{P}^{3/2}_a$~}
\def\AZ{$\mbox{A}^{0}_r$~}
\def\AU{$\mbox{A}^{1}$~}
\def\epsilona{\epsilon^{\scriptscriptstyle{<}}}
\def\epsilonb{\epsilon^{\scriptscriptstyle{>}}}
\def\lgraffa{ \mbox{\Large $\{$ } \hskip -0.2cm}
\def\rgraffa{ \mbox{\Large $\}$ } }
\def\restriction{\upharpoonright}
\def\M{{\scriptscriptstyle{M}}}
\def\m{m}
\def\Fre{Fr\'echet~}
\def\I{{\mathcal N}}
\def\ap{{\scriptscriptstyle{ap}}}
\def\fiap{\varphi_{\ap}}
\def\EEE{ {\textbf E} }
\def\TTT{ {\textbf T} }
\def\KKK{ {\textbf K} }
\def\FFi{ {\bf \Phi} }
\def\GGam{ {\bf \Gamma} }
\def\a{a}
\def\ep{\epsilon}
\def\parn{\par\noindent}
\def\teta{M}
\def\elle{L}
\def\ro{\rho}
\def\si{\sigma}
\def\ga{\gamma}
\def\de{\delta}
\def\la{\lambda}
\def\te{\vartheta}
\def\ch{\chi}
\def\et{\eta}
\def\complessi{\mathbb{C} }
\def\reali{\mathbb{R}}
\def\interi{\mathbb{Z}}
\def\naturali{\mathbb{N}}
\def\bT{{\textbf T}}
\def\T1{{\textbf T}^{1}}
\def\EE{{\mathcal E}}
\def\FF{{\mathcal F}}
\def\GG{{\mathcal G}}
\def\PP{{\mathcal P}}
\def\QQ{{\mathcal Q}}
\def\Np{{\hat{N}}}
\def\Pp{{\hat{P}}}
\def\Pip{{\hat{\Pi}}}
\def\Vp{{\hat{V}}}
\def\Ep{{\hat{E}}}
\def\Fp{{\hat{F}}}
\def\Gp{{\hat{G}}}
\def\Ip{{\hat{I}}}
\def\Tp{{\hat{T}}}
\def\Mp{{\hat{M}}}
\def\La{\Lambda}
\def\Ga{\Gamma}
\def\Si{\Sigma}
\def\Upsi{\Upsilon}
\def\Gag{{\check{\Gamma}}}
\def\Lap{{\hat{\Lambda}}}
\def\Sip{{\hat{\Sigma}}}
\def\Upsig{{\check{\Upsilon}}}
\def\Kg{{\check{K}}}
\def\ellp{{\hat{\ell}}}
\def\j{j}
\def\jp{{\hat{j}}}
\def\Stir{{\mathscr S}}
\def\M{{\mathscr M}}
\def\ess{\mathfrak s}
\def\elg{\mathfrak l}
\def\va{\mathfrak v}
\def\Ma{{\mathfrak M}}
\def\MM{{\mathcal M}}
\def\RR{{\mathcal R}}
\def\BB{{\mathcal B}}
\def\LL{{\mathcal L}}
\def\SS{{\mathcal S}}
\def\DD{{\mathcal D}}
\def\VV{{\mathcal V}}
\def\WW{{\mathcal W}}
\def\OO{{\mathcal O}}
\def\CC{{\mathcal C}}
\def\AA{{\mathcal A}}
\def\CC{{\mathcal C}}
\def\JJ{{\mathcal J}}
\def\NN{{\mathcal N}}
\def\WW{{\mathcal W}}
\def\HH{{\mathcal H}}
\def\XX{{\mathcal X}}
\def\YY{{\mathcal Y}}
\def\ZZ{{\mathcal Z}}
\def\UU{{\mathcal U}}
\def\CC{{\mathcal C}}
\def\XX{{\mathcal X}}
\def\cir{{\scriptscriptstyle \circ}}
\def\circa{\thickapprox}
\def\vain{\rightarrow}
\def\ss{s}
\def\vains{\stackrel{\ss}{\rightarrow}}
\def\parn{\par \noindent}
\def\salto{\vskip 0.2truecm \noindent}
\def\spazio{\vskip 0.5truecm \noindent}
\def\vs1{\vskip 1cm \noindent}
\def\fine{\hfill $\diamond$ \vskip 0.2cm \noindent}
\newcommand{\rref}[1]{(\ref{#1})}
\def\beq{\begin{equation}}
\def\feq{\end{equation}}
\def\beqq{\begin{eqnarray}}
\def\feqq{\end{eqnarray}}
\def\barray{\begin{array}}
\def\farray{\end{array}}
%%%%%%%%% THIS NUMBERS EQUATIONS BY SECTIONS %%%%%%%%%%%%%
\makeatletter \@addtoreset{equation}{section}
\renewcommand{\theequation}{\thesection.\arabic{equation}}
%\thesection instead of \arabic{section} for correct equation numbering
% in appendices
\makeatother
%%%%%%%%%%%%%%%%%%%%%%%%%%%INTESTAZIONE%%%%%%%%%%%%%%%%%%%%%%%%%%%%%%%
\begin{center}
{\huge On a theorem by \Treves \hskip -0.2cm .}
\end{center}
\vspace{0.5truecm}
\begin{center}
{\large
Carlo Morosi${}^1$, Livio Pizzocchero${}^2$} \\
\vspace{0.5truecm} ${}^1$ Dipartimento di Matematica, Politecnico
di
Milano, \\ P.za L. da Vinci 32, I-20133 Milano, Italy \\
e--mail: carmor@mate.polimi.it \\
${}^2$ Dipartimento di Matematica, Universit\`a di Milano\\
Via C. Saldini 50, I-20133 Milano, Italy\\
and Istituto Nazionale di Fisica Nucleare, Sezione di Milano, Italy \\
e--mail: livio.pizzocchero@mat.unimi.it
\end{center}
\vspace{0.3truecm}
{\footnotesize \textbf{Abstract.} According to a theorem in \cite{Tre},
the conserved functionals of the KdV equation vanish on each formal Laurent series
$1/x^2 + u_0 + \sum_{k=2}^{+\infty} u_k x^k$. We propose a new, very simple geometrical
proof for this statement.
\par \vspace{0.2truecm} \noindent \textbf{Keywords:} KdV equation, simmetries and conservation laws,
formal Laurent series.
\par \vspace{0.1truecm} \noindent \textbf{AMS 2000 Subject classifications:} 35Q53, 37K05, 37K10.}
\parn
\section{Introduction.} Three years ago, \Treves obtained a new characterization for the conserved
quantities of KdV theory. Roughly speaking, his result concerns functionals which are integrals of differential
polynomials, and their evaluation on formal Laurent series with complex coefficients
in one variable $x$ (defining the integral as the residue in $x$).
For each functional $h$ of this kind on the Laurent series,
\Treves \cite{Tre} proved the equivalence between a) and b): \parn
a) \textsl{$h$ is a conserved functional for the KdV equation}; \parn
b) \textsl{$h(u)=0$ for each Laurent series of the form $u = 1/x^2 + u_0 +
\sum_{k=2}^{+\infty} u_k x^k$ ($u_0, u_2, u_3,... \in \complessi)$}. \parn
Subsequently, \Treves obtained a similar result for the modified KdV equation and
derived the analogue of a) $\Longrightarrow$ b) for the conserved functionals
of the nonlinear Schr\"odinger equation \cite{Tre2}. \parn
In all cases analysed by \Treves, the
proof of either a) $\Longrightarrow$ b) or b) $\Longrightarrow$ a) is very long.
A simplified derivation for the KdV case, still based on the logic
of the \Treves proof, was given by Dickey; this author also found a new proof of
a) $\Longrightarrow$ b) for the KdV, and established its analogoue for the Boussinesq theory,
using the dressing method for the Lax operator \cite{Dic}. \parn
We became aware of the above results very recently, due to a talk
given in Milano by Prof. \Treves \cite{Tre3}, and we soon developed an interest in a
further simplification of the proofs. We investigated in particular
the implication a) $\Longrightarrow$ b), concentrating for brevity on the KdV case and trying to
isolate \textsl{a single geometrical} \textsl{property} of the conserved functionals,
sufficient to derive the thesis. The conclusion of our analysis is described in this Letter:
here we propose a proof of a) $\Longrightarrow$ b) for the KdV,
different from the ones of \Treves and Dickey and possessing in our opinion the previously asked
feature; the same approach could be probably used for other integrable systems. \parn
Our argument can be described in very few lines, in the following way: \parn
I) the conserved KdV functionals are known to be
invariant under the B\"acklund transformation (often called auto-B\"acklund)
$M \circ R \circ M^{-1}$, where $M$ and $R$ are the Miura and
reflection transformations, respectively. This is the geometrical
property from which everything follows. \parn
II) Any Laurent series $u = 1/x^2 + u_0 +
\sum_{k=2}^{+\infty} u_k x^k$ is the B\"acklund transform of a series
$w = w_0 + \sum_{k=2}^{\infty} w_k x^k$. \parn
III) If $h$ is a conserved KdV functional and $u$, $w$ are as before, we have $h(u) = h(w)$; on the
other hand, $h(w)=0$ for a trivial reason: in fact, this is the integral of a series with no negative
powers of $x$ and thus with zero residue.
The conclusion is $h(u)=0$. \parn
The rest of this Letter is simply a rigorous formulation of items I)-III).
In Section \ref{due}, to fix the language we give some background on differential
polynomials, functionals, KdV theory and state precisely the \Treves theorem; in Section \ref{tre},
we review the B\"acklund transformation and formalize statement I) in the framework of
Laurent series. Expert readers can skip most of the preliminaries in these two Sections, and concentrate on:
Eq.s (\ref{spaceqq}-\ref{intdiff}) describing the space of Laurent series;
Eq.s (\ref{mmrr}-\ref{invab}) on the B\"acklund transformation and the invariance of KdV functionals.
In Section \ref{qua},
we prove II) and show a) $\Rightarrow$ b) along the lines of III). \parn
Let us point out how the idea I)-III) could be employed in relation
to other integrable equations. First of all, one needs a B\"acklund transformation
leaving invariant the conserved functionals. Trivially, all functionals of the theory vanish
on the subspace of formal series with no negative powers of $x$. One should
start from this subspace or a subset of it, and characterize its image
under the B\"acklund transformation; the latter is made of nontrivial Laurent series,
on which the conserved functionals are again zero. In the KdV case,
the starting set and its B\"acklund image consist, respectively, of the series $w$, $u$
mentioned in II). \parn
\textbf{Some terminology.} All vector spaces considered in this Letter are over $\complessi$.
By a differential algebra, we mean an associative and \textsl{commutative} algebra equipped with a derivation,
i.e., with a linear map of the algebra into itself having the Leibnitz property w.r.t. the product. A
morphism of differential algebras is an algebraic morphism respecting the derivations.
\section{Formal variational calculus, KdV and the \Treves theorem.}
\label{due}
In all concrete manipulations,
the KdV equation $(d/ d t) q = q_{x x x} - 12 q q_{x}$ is understood
as a vector field on some "space" $\QQ$, whose elements $q$
are "functions of one variable $x$". The analysis of this vector field
is greatly simplified if one assumes $\QQ$ to be closed
under pointwise sums and products, and under the operation
$q \mapsto q_{x}$ of derivation w.r.t. $x$; in this case, $\QQ$ is a
differential algebra. \parn
Investigations in this area soon made clear that
the striking features of KdV are largely independent of the choice
of the differential algebra $\QQ$; the same can be said for other integrable
PDEs, discovered shortly after it. To take this fact into account,
Gelfand and Dickey (see \cite{GeD} and references therein) invented
a \textsl{formal variational calculus}, allowing to describe
the KdV and similar systems within a very pure algebraic setting.
Hereafter we illustrate some facts about this calculus, in a fashion convenient for our
purposes (and partly inspired by the setting of \cite{Tre}). \parn
Formal variational calculus for  KdV theory
can be based on the commutative algebra
\beq \AAA := \complessi[\gen, \gen_{x}, \gen_{xx},...]_{0}~, \feq
made of complex polynomials in infinitely many
indeterminates $\gen, \gen_x$,$\gen_{xx}$, ... without constant term.
$\AAA$ becomes a differential algebra, when equipped with the unique derivation
$\DDD$ such that ({\footnote{One occasionally needs the full algebra
$\complessi[\gen, \gen_{x}, \gen_{xx},...]$, including
polynomials with constant term. This is an algebra
with unity, containing $\AAA$ as an ideal and identifiable with
$\AAA \oplus \complessi$ as a vector space;  the derivation $\DDD$ is extended to this larger
algebra setting $\DDD 1 := 0$. However, this enlargement plays no role in our construction.}})
\beq \DDD \gen = \gen_{x},\qquad \DDD \gen_{x} = \gen_{xx}, \qquad ...~. \feq
We write $\FFF,\GGG$, etc. for the elements of $\AAA$, and $\FFF \GGG$
for their product as polynomials. The \textsl{composition product} $\FFF \ccc \GGG \in \AAA$
is the polynomial obtained from the expression of $\FFF$ replacing $\gen, \gen_x,...$ with
$\GGG$, $\DDD \GGG,...$
({\footnote{For example, if $\FFF = \gen + \gen^2_{x}$ and
$\GGG = \gen^4 + \gen_{xx}$
we have: $\FFF \GGG =$ $\gen^5 + \gen^4 \gen^2_{x} + \gen \gen_{xx} + \gen^2_{x} \gen_{xx}$;
$\FFF \ccc \GGG =$ $\GGG + (\DDD \GGG)^2 =$ $(\gen^4 + \gen_{x x}) + (4 \gen^3 \gen_x + \gen_{xxx})^2 =
\gen^4 + 16 \gen^6 \gen^2_x + \gen_{xx} + 8 \gen^3 \gen_x \gen_{xxx} + \gen^2_{xxx}$.}}).
For each fixed $\GGG$, the mapping $\FFF \mapsto \FFF \ccc \GGG$ is the unique automorphism
of the differential algebra $\AAA$ sending $\gen$ into $\GGG$. The operation $\ccc$
is associative, so $(\AAA, \ccc)$ is a monoid with unit $\gen$.
\parn
Let us consider any differential algebra $(\QQ, \boma{\cdot}_x)$ (of elements $q, p,...$,
with a derivation $q \in \QQ \mapsto q_x \in \QQ$; this notation for the derivation is purely
conventional). Then, we can represent the elements of $\AAA$ as
\textsl{transformations of $\QQ$} into itself. More precisely, if $\FFF \in \AAA$ and
$q \in \QQ$, let us denote with $F(q) \in \QQ$ the element obtained from the
expression of $\FFF$ substituting the symbols $\gen$, $\gen_x$, ... with $q$, $q_x$, etc.~.
In this way, $\FFF$ induces a map of polynomial type
({\footnote{A map $K : \SS \vain \TT$, where $\SS$ and $\TT$ are vector spaces, is said to be
of polynomial type if there are $m$-linear maps $K_m : \times^m \SS \vain \TT$
$(m=0,...,n)$ such that $K(s) = \sum_{m=0}^{n} K_m(s,...,s)$ for all $s \in \SS$. We will write
$Pol(\SS,\TT)$ for the maps of polynomial type between $\SS$ and $\TT$.}})
\beq F : \QQ \vain \QQ~, \qquad q \mapsto F(q)~. \feq
We point out the remotion of bold typeface to distinguish
this map from $\FFF$; in particular, the transformation $\xi : \QQ \vain \QQ$ induced by $\gen$ is
just the identity map $q \mapsto q$. As $\FFF$ ranges over the whole $\AAA$, we get a correspondence
\beq \FFF \vain Pol(\QQ,\QQ)~, \qquad \FFF \mapsto F~. \label{weget} \feq
Now, the set $Pol(\QQ,\QQ)$ of polynomial maps $\QQ \vain \QQ$ is itself a
commutative algebra, with all the operations defined pointwisely: for $K, L : \QQ \vain \QQ$
and $\lambda \in \complessi$,
$K + L$, $\lambda K$, $K L : \QQ \vain \QQ$ are the maps $q \mapsto K(q) + L(q)$, $q \mapsto \lambda K(q)$,
$q \mapsto K(q) L(q)$.
Furthermore, $Pol(\QQ,\QQ)$ becomes a differential algebra with the derivation $\partial : K \mapsto \partial K$
such that $(\partial K)(q) := K(q)_x$ for all $q \in \QQ$. One easily recognizes that \rref{weget} is
a morphism of differential algebras: for all $\FFF, \GGG \in \AAA$ and $\lambda \in
\complessi$, the transformations
corresponding to $\FFF + \GGG$, $\lambda \FFF$, $\FFF \GGG$, $\DDD \FFF$ are $F + G$, $\lambda F$,
$F G$, $\partial F$. \parn
$Pol(\QQ,\QQ)$ is also a monoid with the usual composition of maps
$F \circ G : q \mapsto F(G(q))$ and the identity map as unit;
it turns out that \rref{weget} is a monoid morphism between
$(\AAA, \ccc)$ and $(Pol(\QQ,\QQ), \circ)$.
\parn
Due to the previous facts, it is helpful for intuition \textsl{to think the elements of $\AAA$
as transformations, even when no differential algebra $(\QQ, \boma{\cdot}_x)$ is specified}. \parn
The next step
in formal variational calculus is the introduction of \textsl{functionals},
which are "integrals" of transformations. The only property needed for the integral is
to vanish on a derivative; for this reason Gelfand and Dickey defined this operation as the quotient map
\beq \iint : \AAA \vain \AAA/\Im \DDD~, \feq
and called functionals the elements of $\AAA/\Im \DDD$; each of them has
the form
\beq \fff = \iint \hskip 0.02cm \FFF \qquad (\FFF \in \AAA)~. \feq
For any
"transformation" $\GGG \in \AAA$, the functional
\beq \fff \ccc \GGG := \iint  (\FFF \ccc \GGG) \in \AAA/\Im \DDD \feq
is well defined (i.e., independent on the choice of $\FFF$ within the equivalence class $\fff$);
we call this the \textsl{composition} between $\fff$ and $\GGG$. One easily checks the
associative property $(\fff \ccc \GGG) \ccc \GGG' = \fff \ccc (\GGG \ccc \GGG')$
for any $\GGG' \in \AAA$. \parn
To get concrete counterparts of functionals, consider any differential algebra $(\QQ,\cdot_{x})$, and
define an \textsl{integration} for it to be any linear map
\beq \mbox{$\int$} : \QQ \vain \complessi~~\mbox{such that $\int q_x = 0~~\forall q \in \QQ$}~; \feq
the triple $(\QQ, \boma{\cdot}_x, \int)$ will then be called an \textsl{integral-differential} algebra.
If $\fff = \iint \FFF \in \AAA/\Im \DDD$, define
\beq f : \QQ \vain \complessi~, \qquad q \mapsto f(q) := \mbox{$\int$} F(q)~; \label{pol} \feq
this definition is well posed, and gives a linear correspondence
\beq \AAA/\Im \DDD \vain Pol(\QQ,\complessi)~, \qquad \fff \mapsto f~; \feq
for all $\fff$ as above and $\GGG \in \AAA$, the map $\QQ \vain \complessi$ induced
by $\fff \ccc \GGG$ is the usual composition $f \circ G: q \mapsto f(G(q))$. \parn
One can then go on at the level of $\AAA$, defining notions such as
\textsl{vector fields} (identifiable with elements of $\AAA$), and the (Lie)
\textsl{derivative} of a functional $\hhh$ along a vector field $\XXX$; if the latter vanishes,
we say that $\hhh$ \textsl{is conserved by}  $\XXX$ (see \cite{GeD} and references therein). \parn
All this machinery is designed to discuss topics such as
the \textsl{KdV vector field} and its conserved functionals, i.e.,
\beq \XXX_{\KdV} := \gen_{xxx} - 12 \gen \gen_x \feq
\beq \ZZZ_{\KdV} := \{ \hhh \in \AAA/Im \DDD~|~\hhh~\mbox{is a conserved by $\XXX_{\KdV}$} \}~.
\feq
An outstanding feature of KdV theory is that $\ZZZ_{\KdV}$ is \textsl{infinite dimensional}
(as a vector space over $\complessi$). A basis for it is well known and consists of
countably many functionals $(\hhh_k)_{k =1,2,...}$, for which several equivalent
constructions are available: for example, one can use the Magri-Lenard recursion scheme
\cite{Mag}. The first elements are
\beq \hhh_1 := - {1 \over 4} \iint \hskip 0.02cm \gen,~~\hhh_2 := {1 \over 2} \iint \, \hskip -0.00cm \gen^2,~~
\hhh_3 := - \, \iint \hskip 0.01cm (2 \gen^3 + {1 \over 2} \gen_x^2),~~\hhh_4 :=
\iint \hskip 0.02cm (10 \gen^4 + 10 \gen \gen_x^2 + {1 \over 2} \gen_{xx}^2)~.
\label{hk} \feq
We finally come to the \Treves theorem, concerning the KdV
conserved functionals an their representation on a particular integral-differential
algebra $(\QQ, \boma{\cdot}_x, \int)$. By definition, this is made of formal
Laurent series in one indeterminate $x$ and complex coefficients, i.e.
\beq \QQ := \{ q = \sum_{k} q_k x^k~|~k \in \interi,~q_k \in \complessi~\forall k,~
q_k = 0~\mbox{for $k \leq k_{*} = k_{*}(q)$} \}~; \label{spaceqq} \feq
$\QQ$ is a commututative algebra with usual Cauchy product; it carries the derivation and
integration
\beq \boma{\cdot}_x : \QQ \vain \QQ, \quad q \mapsto q_{x} := \sum_{k} k q_k x^{k-1}~;
\qquad \mbox{$\int$} : \QQ \vain \complessi, \quad q \mapsto \mbox{$\int$} q := q_{-1}~. \label{intdiff} \feq
Clearly, $\int q = 0$ iff $q = p_x$ for some $p \in \QQ$; of course, the definition of $\mbox{$\int q$}$
as the "residue" $q_{-1}$ suggests to interpret it as a loop integral about zero. \parn
With the previous notations, the \Treves  theorem reads:
\begin{prop}
\label{teot}
\textbf{Proposition} \cite{Tre}. For any $\hhh \in \AAA/\Im \DDD$, statements a) and b) are
equivalent: \parn
a) $\hhh \in \ZZZ_{\KdV}$; \parn
b) it is $h(u)=0$ for each $u \in \QQ$ of the form $u = 1/x^2 + u_0 +
\sum_{k=2}^{+\infty} u_k x^k$.
\end{prop}
As anticipated, the rest of this Letter is a new geometrical proof of the implication  $a) \Longrightarrow b)$.
\parn
\section{A review of the Miura and B\"acklund transformations.}
\label{tre}
The basic facts on these transformations can be stated in the language of
formal variational calculus; so, we
consider the algebra $\AAA$ of the previous Section, and state the following
\begin{prop}
\textbf{Definition.} The \textsl{Miura and reflection transformations} are
\beq \MMM := \gen_x + 2 \gen^2~; \qquad \RRR := -\gen~. \feq
\end{prop}
Both $\MMM,\RRR$ are elements of $\AAA$, so they can be composed as explained
previously; of course $\MMM \ccc \RRR= - \gen_x + 2 \gen^2$. We can as well compose
functionals $\in \AAA/\Im \DDD$ with these transformations; for example, composing the
first KdV conserved functionals \rref{hk} with the Miura transformation we obtain
\beq \hhh_1 \ccc \MMM = - {1 \over 2} \iint \gen^2,~~\hhh_2 \ccc \MMM
= \iint (2 \gen^4 + {1 \over 2} \gen_x^2),~~
\hhh_3 \ccc \MMM = - \iint (16 \gen^6 + 20 \gen^2 \gen_x^2 + {1 \over 2} \gen_{xx}^2),~~ \label{mhk}\feq
$$ \hhh_4 \ccc \MMM = \iint( 160 \gen^8 + 560 \gen^4 \gen_{x }^2 + 18 \gen_{x }^4
+ 96 \gen\gen_{x }^2 \gen_{x x} + 28 \gen^2\gen_{x x}^2 + {1 \over 2} \gen_{x x x}^2)~. $$
The following facts are known from the very beginning of KdV history: \parn
\begin{prop}
\label{miuma}
\textbf{Proposition.} For each $\hhh \in \ZZZ_{\KdV}$: \parn
i) $\hhh \ccc \MMM$ is a conserved functional for the
\textsl{modified KdV vector field} $\XXX_{\mKdV} := \gen_{xxx} - 24 \, \gen^2 \gen_{x}$. \parn
ii) $\hhh \ccc \MMM$ is invariant under reflection:
$(\hhh \circ \MMM) \circ \RRR = \hhh \circ \MMM$.
\end{prop}
\textbf{References for the proof.} For i), see the original papers by Miura \textsl{et al} \cite{Miu}. ii)
is proved recursively for all elements $(\hhh_k)_{k=1,2,...}$ in the basis of $\ZZZ_{\KdV}$, using
the Magri-Lenard recursion relations connecting
$\hhh_k \circ \MMM$ to $\hhh_{k+1} \circ \MMM$ \cite{Mag}: these relations are reflection invariant. \fine
\parn
For our purposes, ii) is the essential feature of $\MMM$ and $\RRR$; now we represent this result
on any integral-differential algebra $(\QQ,\boma{\cdot}_x, \int)$. Let us consider
the maps of $\QQ$ into itself induced by $\MMM, \RRR$ according to the framework of the previous
Section; these are
\beq M: \QQ \vain \QQ, ~~p \mapsto M(p) = p_{x} + 2 p^2~; \qquad
\qquad R : \QQ \vain \QQ, ~p \mapsto R(p) = -p \label{mmrr} \feq
(the letter $p$ for elements of $\QQ$ is used here for future conveniency). The above maps will be called the
\textsl{Miura and reflexion transformations on} $\QQ$. Let us also recall that any functional $\fff \in \AAA/\Im \DDD$
induces a map $f : \QQ \vain \complessi$; in particular, considering the KdV conserved functionals
we infer from Prop. \ref{miuma} ii) that
\beq (h \circ M) \circ R = h \circ M \qquad \mbox{for each $\hhh \in \ZZZ_{\KdV}$}~, \label{invar} \feq
with $\circ$ the usual composition of maps. \parn
We go on and introduce the B\"acklund transformation on $\QQ$; essentially, this is the composition of maps
$M \circ R \circ M^{-1}$, leaving invariant any conserved KdV functional due to Eq. \rref{invar}.
However, $M$ is typically non invertible on the full space $\QQ$: to overcome this difficulty, we use
the following
\begin{prop}
\textbf{Definition.} Consider the set $2^{\QQ}$ of the parts of $\QQ$ (i.e., the collection of
all subsets of $\QQ$). The \textsl{B\"acklund transformation} on $\QQ$ is the set-valued map
\beq B : \QQ \vain 2^{\QQ}, \quad q \mapsto B(q) := \{ (M \circ R)(p) ~|~p \in \QQ~, M(p) = q~\}~.
\label{setv} \feq
\end{prop}
With this definition, Eq. \rref{invar} implies
\begin{prop}
\label{corinvab}
\textbf{Proposition.} Let $\hhh \in \ZZZ_{\KdV}$; then the map $h : \QQ \vain \reali$ is B\"acklund
invariant, in the following sense: for all $q, r \in \QQ$,
\beq r \in B(q) \quad \Longrightarrow \quad h(r) = h(q)~. \label{invab} \feq
\end{prop}
\section{The implication a) $\boma{\Longrightarrow}$ b) in Prop. \ref{teot}: a new proof.}
\label{qua}
The simple geometrical proof we propose is based on the scheme I)-III) of the Introduction. Item I)
has been treated in the previous Section; here we formalise II) III). From now on,
$(\QQ, \boma{\cdot}_x, \int)$ is the integral-differential algebra
(\ref{spaceqq}-\ref{intdiff}) of formal Laurent series.
\begin{prop}
\textbf{Definition.} We put
$$ \WW := \{ w \in \QQ~|~ w = w_0 + \sum_{k=2}^{+\infty} w_k x^k~\}~; \qquad
\VV := \{ v \in \QQ~|~ v = {1 \over 2 x} + v_1 x + \sum_{k=3}^{+\infty} v_k x^k~\}~; $$
\beq \UU := \{ u \in \QQ~|~ u = {1 \over x^2} + u_0 + \sum_{k=2}^{+\infty} u_k x^k~\}~. \label{uvw} \feq
\end{prop}
\begin{prop}
\textbf{Lemma.} Consider the Miura and reflections transformations $M, R$ of Eq. \rref{mmrr}. Then: \parn
i) $M$ is one to one between $\VV$ and $\WW$; \parn
ii) $M \circ R$ is one to one between $\VV$ and $\UU$. \parn
\end{prop}
\textbf{Proof.} i) For all $v \in \VV$, an elementary computation gives
$$ M(v) = v_x + 2 v^2 =
3 v_1 + (5 v_3 + 2 v_1^2) x^2 + 6 v_4 x^3 + (7 v_5 + 4 v_1 v_3) x^4 + (8 v_6 + 4 v_1 v_4) x^5 + $$
\beq + \sum_{k=6}^{\infty} \Big( (k+3) v_{k+1} + 4 v_1 v_{k-1} + 2 \sum_{j=3}^{k-3} v_{j} v_{k-j} \Big) x^k~; \feq
from here, we see that $M(v) \in \WW$. Now, let us consider any $w \in \WW$ and show that the equation
$M(v) = w$ has a unique solution $v \in \VV$. In fact, $M (v) = w$ is equivalent to
$3 v_1 = w_0$, $5 v_3 + 2 v_1^2 = w_2$, etc., giving
$$ v_1 = {1 \over 3} w_0,~, \qquad v_3 = - {2 \over 5} v_1^2 + {1 \over 5} w_2 = -{2 \over 45} w_0^2  +
{1 \over 5} w_2~, ~~ v_4 = {1 \over 6} w_3~, $$
\beq v_5 =  {8 \over 945} w_0^3 - {4 \over 105} w_0 w_2 + {1 \over 7} w_4~, \quad
v_6 = -{1 \over 36} w_0 w_3 + {1 \over 8} w_5~, \feq
$$ v_{k+1} = - {2 \over k + 3} \Big( 2 v_1 v_{k-1} + \sum_{j=3}^{k-3} v_j v_{k-j} \Big) + {w_k \over k+3}
\qquad \mbox{for all $k \geqs 6$}~; $$
the equation in the last line is a recursion formula, determining uniquely $v_k$ for all
$k \geqs 7$. \parn
ii) For all $v \in \VV$, we find
$$ (M \circ R)(v) = - v_x + 2 v^2 =
{1 \over x^2} + v_1 + (- v_3 + 2 v_1^2) x^2 - 2 v_4 x^3 + (-3 v_5 + 4 v_1 v_3) x^4 + (- 4 v_6 + 4 v_1 v_4) x^5 + $$
\beq + \sum_{k=6}^{\infty} \Big( (1-k) v_{k+1} + 4 v_1 v_{k-1} + 2 \sum_{j=3}^{k-3} v_{j} v_{k-j} \Big) x^k~; \feq
this shows that $(M \circ R)(v) \in \UU$. For all $u \in \UU$, the equation $(M \circ R)(v) = u$ has a unique
solution $v \in \VV$, given by
$$ v_1 = u_0,~, \qquad v_3 = 2 u_0^2 - u_2~, ~~ v_4 = - {1 \over 2} u_3~, $$
\beq v_5 =  {8 \over 3} u_0^3 - {4 \over 3} u_0 u_2 - {1 \over 3} u_4~, \quad
v_6 = -{1 \over 2} u_0 u_3 - {1 \over 4} u_5~, \feq
$$ v_{k+1} = {2 \over k - 1} \Big( 2 v_1 v_{k-1} + \sum_{j=3}^{k-3} v_j v_{k-j} \Big) - {u_k \over k-1}
\qquad \mbox{for all $k \geqs 6$}~. $$
\fine
Of course, the previous Lemma implies:
\begin{prop}
\textbf{Corollary.} Define a \textsl{restricted B\"acklund transformation}
\beq B_0 : \WW \vain \UU~, \qquad B_0 := ((M \circ R) \restriction \VV) \circ (M \restriction \VV)^{-1}~; \feq
then, $B_0$ is one to one between $\WW$ and $\UU$.
\end{prop}
The final step in our argument is trivial:
\begin{prop}
\label{holom}
\textbf{Lemma.} Consider any functional $\fff \in \AAA/\Im \DDD$; then $f$ vanishes on the
"holomorphic subspace"
\beq \ZZ := \{ z \in \QQ~|~z = \sum_{k=0}^{\infty} z_k x^k~\}~. \feq
\end{prop}
\textbf{Proof.} $\ZZ$ is a differential subalgebra of $\QQ$, and
$\int$ clearly vanishes on $\ZZ$. Consider any functional
$\fff = \, \iint \hskip 0.01cm \FFF$. For all $z \in \ZZ$ we have $F(z) \in \ZZ$ and
$f(z) = \int F(z)=0$. \fine
We are finally ready to give our
\vskip 0.1cm \noindent
\textbf{Proof of the implication a) $\boma{\Longrightarrow}$ b) in the \Treves theorem.} Consider
a functional $\hhh \in \ZZZ_{\KdV}$, and any Laurent series $u \in \UU$. By the previous Corollary,
there is a unique $w \in \WW$ such that $u = B_0(w)$; of course this implies $u \in B(w)$, with
$B$ the (set-valued) B\"acklund transformation \rref{setv}. These facts give
\beq h(u) = h(w) = 0~. \feq
The first equality above is ensured by the B\"acklund invariance of $h$ (Prop. \ref{corinvab}); the
second one follows from Lemma \ref{holom} and the evident inclusion $\WW \subset \ZZ$. \fine
\vskip 0.6cm \noindent
\textbf{Acknowledgments.} This work was partly supported by INDAM, Gruppo Nazionale per la
Fisica Matematica.

\end{document}